# ON THE QUADRATIC INTERACTION TERM IN THE REAL DIRAC FIELD THEORY

S.Botrić* and K.Ljolje
Academy of Sciences and Arts of Bosnia and Herzegovina,
Sarajevo,Bosnia and Herzegovina

Summary.-Effects of the quadratic interaction term in the real Dirac field theory are considered.

* Faculty of Electrical Engineering,Mechanical Engineering and Naval Architecture,University of Split,Split,Croatia

## 1. INTRODUCTION

The Dirac field theory in real domain,analogously to the Maxwell-Lorentz electromagnetic field theory,has been developed [1,2]. Reduction of the self-interaction in linear approximation of interaction with the electromagnetic field has been analysed in the article [3].There has been shown that the self-interaction can be reduced analogously to the classical field theory [4-7].Application to the hydrogen atom yields energies of stationary states exactly equal to those of the conventional relativistic theory.By this internal and observable consistence of the real Dirac field theory has been established on the linear level.The question arises what is situation on higher levels.We investigate this question with respect to the quadratic term in this article.

## 2. LAGRANGIAN WITH QUADRATIC INTERACTION TERM

The Lagrangian density of the real Dirac field interacting with the electromagnetic field [1] in approximation up to the quadratic interaction term with respect to $A^{\alpha}$ is given by

$$L_{DI}^{(2)} = L_{DI}^{(1)} + \frac{e^2}{K} A_{\beta} A^{\beta} \overline{\Psi}\Psi \tag{1}$$

where

$$L_{DI}^{(1)} = K(\overline{\Psi}\Psi - \kappa^2 \overline{\Phi}\Phi) - \frac{1}{c} A_{\alpha} j_D^{\alpha} \quad , \quad K = mc^2 \tag{2}$$

$$\frac{1}{c} j_D^{\alpha} = e(\overline{\Psi}\eta^{\alpha}\Psi + \kappa^2 \overline{\Phi}\eta^{\alpha}\Phi) \tag{3}$$

It is necessary to reduce the self-interaction now in this Lagrangian.The corresponding situation doesn't exist in the classical field theory.Consequently,there is not prescription for such a reduction.There is only requirement that in absence of external field there exist solutions which correspond to a free particle motion.

It is reasonable to make comparison of the equation of the real Dirac field with full self-interaction

$$D \frac{1-a}{1-a^2} \Psi + \kappa^2 (1+a)\Phi = 0 \quad , \quad a = \frac{e}{K} A_{\beta}\eta^{\beta} \tag{4}$$

with desirable equation

$$D \frac{1-(a-b)}{1-(a-b)^2} \Psi + \kappa^2 (1+a-b)\Phi = 0 \tag{5}$$

which in the quadratic approximation becomes

$$D[1 - a + a^2 + (b - ab - ba + b^2)]\Psi + \kappa^2(1 + a - b)\Phi = 0 \tag{6}$$

where $ab + ba = 2\frac{e^2}{K^2}A_\alpha B^\alpha$.

Relying on this comparison we define

$$L_s = \frac{1}{2c}A_{\text{ret}\alpha}j_D^\alpha + \frac{e^2}{K}A_{\text{Dret}\alpha}A_{\text{Dret}}^\alpha\overline{\Psi}\Psi - 2\frac{e^2}{K}A_\alpha A_{\text{Dret}}^\alpha\overline{\Psi}\Psi \tag{7}$$

where

$$A_{\text{Dret}}^\alpha = \int\frac{1}{|\vec{r} - \vec{r}'|}\frac{1}{c}j_{\text{Dret}}^\alpha d^3x' \tag{8}$$

The new Lagrangian is then given by

$$L = L_{em} + L_{DI}^{(2)} + L_s \tag{9}$$

In the introduction we have mentioned that we are going to perform calculation up to the terms of the order $\alpha^4$.

## 3. EQUATIONS OF MOTION

Starting with the variational principle

$$\delta S = 0 \quad , \quad S = \int L\frac{1}{c}d^4x \tag{10}$$

one obtains the Lagrange's equations for $A^\alpha$ and $\Phi$:

$$\partial_\mu\partial^\mu A^\alpha = \frac{4\pi}{c}j_D^\alpha + 4\pi\frac{2e^2}{K}(-A^\alpha + A_{\text{Dret}}^\alpha)\overline{\Psi}\Psi \tag{11}$$

$$D\left[1 - a + b + 2\frac{e^2}{K^2}\int\frac{1}{|\vec{r} - \vec{r}'|}\left((A_{\text{Dret}\alpha} - A_\alpha)\overline{\Psi}\Psi\right)_{\text{adv}}\eta^\alpha d^3x' + \right.$$

$$\left. + \frac{e^2}{K^2}(A_{\text{Dret}\alpha} - A_\alpha)(A_{\text{Dret}}^\alpha - A^\alpha)\right]\Psi +$$

$$+ \kappa^2\left[1 + a - b - 2\frac{e^2}{K^2}\int\frac{1}{|\vec{r} - \vec{r}'|}\left((A_{\text{Dret}\alpha} - A_\alpha)\overline{\Psi}\Psi\right)_{\text{adv}}\eta^\alpha d^3x'\right]\Phi = 0 \tag{12}$$

A particular solution of the equation (11) is given by

$$A_p^\alpha = A_{\text{Dret}}^\alpha \tag{13}$$

After substitution of (13) into (12) one gets

$$D(1 - a_{\text{rad}})\Psi + \kappa^2(1 + a_{\text{rad}})\Phi = 0 \tag{14}$$

where $a_{rad} = a - b$.

Therefore,the reduction of the self-interaction by (7) leads to radiation effects in agreement with expressed criteria.

## 4. HYDROGEN ATOM

We apply the procedure from Section 3 to the hydrogen atom.Let the proton be at rest

$$\frac{1}{c} j^{\alpha}_{ext} = (|e|\delta(\vec{r}), 0) \tag{15}$$

The Lagrangian density (9) gets the additional term

$$L_{ext} = -\frac{1}{c} A_{\alpha} j^{\alpha}_{ext} \tag{16}$$

This term changes the equation (11) into

$$\partial_{\mu}\partial^{\mu} A^{\alpha} = \frac{4\pi}{c}\left(j^{\alpha}_{D} + j^{\alpha}_{ext}\right) + 4\pi 2\frac{e^2}{K}\left(-A^{\alpha} + A^{\alpha}_{Dret}\right)\overline{\Psi}\Psi \tag{17}$$

The new particular solution is given by

$$A^{\alpha}_{p} = A^{\alpha}_{Dret} + A^{\alpha}_{extret} \tag{18}$$

Substitution of (18) into (12) yields

$$D\left[1 - a^{(0)}_{ext} + a^{(o)^2}_{ext} - a_{rad} + 2\frac{e^2}{K^2}\int \frac{e}{|\vec{r} - \vec{r}'|}\left[\left(A^{(o)}_{extret\alpha}\overline{\Psi}\Psi\right)_{ret} - \right.\right.$$

$$\left.\left. - \left(A^{(o)}_{extret\alpha}\overline{\Psi}\Psi\right)_{adv}\right]\eta^{\alpha} d^3x'\right]\Psi +$$

$$+ \kappa^2\left[1 + a^{(o)}_{ext} + a_{rad} - 2\frac{e^2}{K^2}\int \frac{e}{|\vec{r} - \vec{r}'|}\left[\left(A^{(o)}_{extret\alpha}\overline{\Psi}\Psi\right)_{ret} - \left(A^{(o)}_{extret\alpha}\overline{\Psi}\Psi\right)_{adv}\right]\eta^{\alpha} d^3x'\right]\Phi = 0 \tag{19}$$

$$A^{\alpha(o)}_{extret} = \int \frac{1}{|\vec{r} - \vec{r}'|}\frac{1}{c} j^{\alpha}_{extret} d^3x'$$

The expression $1 - a^{(o)}_{ext} + a^{(o)^2}_{ext}$ we may write in the form

$$1 - a^{(o)}_{ext} + a^{(o)^2}_{ext} = \frac{1 - a^{(o)}_{ext}}{1 - a^{(o)^2}_{ext}} \tag{20}$$

(keeping the terms up to $\alpha^4$ order).

The terms besides $a^{(o)}_{ext}$ are generally small.The perturbation method may be applied

$$\Phi = \Phi^{(o)} + \Phi^{(1)} + \ldots \tag{21}$$

with

$$D\frac{1-a_{ext}^{(o)}}{1-a_{ext}^{(o)^2}}\Psi^{(o)} + \kappa^2\left(1+a_{ext}^{(o)}\right)\Phi^{(o)} = 0\,, \tag{22}$$

$$D\frac{1-a_{ext}^{(o)}}{1-a_{ext}^{(o)^2}}\Psi^{(1)} + \kappa^2\left(1+a_{ext}^{(o)}\right)\Phi^{(1)} +$$

$$+\left(-a_{rad}^{(o)} + 2\frac{e^2}{K^2}\int\frac{e}{|\vec{r}-\vec{r}'|}\left[\left(A_{extret\alpha}^{(o)}\overline{\Psi}^{(o)}\Psi^{(o)}\right)_{ret} - \left(A_{extret\alpha}^{(o)}\overline{\Psi}^{(o)}\Psi^{(o)}\right)_{adv}\right]\eta^{\alpha}d^3x'\right)\Psi^{(o)} +$$

$$+\kappa^2\left[a_{rad}^{(o)} - 2\frac{e^2}{K^2}\int\frac{e}{|\vec{r}-\vec{r}'|}\left[\left(A_{extret\alpha}^{(o)}\overline{\Psi}^{(o)}\Psi^{(o)}\right)_{ret} - \left(A_{extret\alpha}^{(o)}\overline{\Psi}^{(o)}\Psi^{(o)}\right)_{adv}\right]\eta^{\alpha}d^3x'\right]\Phi^{(o)} = 0, \tag{23}$$

.

.

.

In accordance to [1-3] we investigate the family of solutions

$$\Psi^{(o)} = \kappa\left(1+a_{ext}^{(o)}\right)N\Phi^{(o)} \tag{24}$$

$$D\left[1-\kappa\left(1+a_{ext}^{(o)}\right)N\right]\Phi^{(o)} = 0 \tag{25}$$

S-transformation of (25) yields

$$\Phi^{(o)'} = S\Phi^{(o)} \equiv \begin{bmatrix}\varphi_a^{(o)}\\ \varphi_b^{(o)}\end{bmatrix} \quad , \quad \varphi_b^{(o)} = N_b\varphi_a^{(o)*} \tag{26}$$

$$\left[i\partial_\alpha\gamma^\alpha - \kappa\left(1+\tilde{a}_{ext}^{(o)}\right)\right]\varphi_a^{(o)} = 0 \quad , \quad \tilde{a}_{ext}^{(o)} = \frac{e}{K}A_{extret\alpha}^{(o)}\gamma^\alpha \tag{27}$$

The equation (27) is standard Dirac equation.A stationary state solution is given by

$$\varphi_a^{(o)} = \varphi_{l\gamma}(\vec{r})e^{-ik_{ol}x^o} \tag{28}$$

Substitution of (28) into (23) yields

$$a_{rad}^{(o)} = 0 \quad , \quad \left(A_{extret\alpha}^{(o)}\overline{\Psi}^{(o)}\Psi^{(o)}\right)_{ret} = \left(A_{extret\alpha}^{(o)}\overline{\Psi}^{(o)}\Psi^{(o)}\right)_{adv} \quad ,$$

and $\Phi^{(1)} = 0$.From here we conclude that stationary solutions of the real Dirac field equation up to the first order of the perturbation method are consistent with the theory of Dirac equation,but the corresponding real Dirac field function is given by $\Phi$.For example

$$\Phi_{2s1/2} = \frac{1}{\kappa\sqrt{8\pi}} \begin{bmatrix} -g_{1-1}\cos k_{02}x^{o} \\ -g_{1-1}\sin k_{02}x^{o} \\ 0 \\ 0 \\ f_{1-1}\cos\vartheta\cos k_{02}x^{o} \\ f_{1-1}\cos\vartheta\sin k_{02}x^{o} \\ -f_{1-1}\sin\vartheta\cos(k_{02}x^{o}-\varphi) \\ +f_{1-1}\sin\vartheta\sin(k_{02}x^{o}-\varphi) \end{bmatrix} \qquad (29)$$

This result we find as further support of the real Dirac field theory,that means of the Dirac field as a classical field analogous to the electromagnetic field.Transition processes we consider elsewhere.

## 4. CONCLUSION

The presented results show that real Dirac field theory is internally and observably consistent also on the quadratic interaction level.